\documentclass[letter]{aa} 
\usepackage{aas_macros}
\usepackage{graphicx}
\usepackage{txfonts}
\usepackage{hyperref}
\graphicspath{{Images/}{../Images/}}
\usepackage{subfiles}
\usepackage{xcolor}
\usepackage[normalem]{ulem}

\begin{document}

    \title{On the hypothesis of an inverted Z-gradient  inside Jupiter}

    \author{S. Howard \inst{1}
          \and T. Guillot\inst{1}
          \and S. Markham\inst{1}
          \and R. Helled\inst{2}
          \and S. Müller\inst{2}
          \and D. J. Stevenson\inst{3}
          \and J. I. Lunine\inst{4,5}
          \and Y. Miguel\inst{6,7}
          \and N. Nettelmann\inst{2}
          }

    \institute{Université Côte d'Azur, Observatoire de la Côte d'Azur, CNRS, Laboratoire Lagrange, France\\
              \email{saburo.howard@oca.eu}
        \and
             Institute for Computational Science, Center for Theoretical Astrophysics \& Cosmology, University of Zurich, Winterthurerstr. 190, CH8057 Zurich, Switzerland,
        \and
             Division of Geological and Planetary Sciences, California Institute of Technology, Pasadena, California 91125, USA
        \and Department of Astronomy, Cornell       University, 122 Sciences Drive,        Ithaca, NY 14853, USA
        \and Carl Sagan Institute, Cornell          University, 122 Sciences Drive,        Ithaca, NY 14853, USA
        \and SRON Netherlands Institute for         Space Research, Niels Bohrweg 4,       2333 CA Leiden, The Netherlands
        \and Leiden Observatory, University of      Leiden, Niels Bohrweg 2, 2333 CA       Leiden, The Netherlands}
    \date{Submitted Oct 2, 2023, Accepted Nov 3, 2023}
    
 
  \abstract
   {Models of Jupiter's interior struggle to agree with measurements of the atmospheric composition. Interior models favour a subsolar or solar abundance of heavy elements $Z$ while atmospheric measurements suggest a supersolar abundance. One potential solution may be the presence of an inverted Z-gradient, namely an inward decrease of $Z$, which implies a larger heavy element abundance in the atmosphere than in the outer envelope.
   }
   {We investigate two scenarios in which the inverted $Z$ gradient is located either where helium rain occurs ($\sim \,$Mbar level) or at upper levels ($\sim \,$kbar level) where a radiative region could exist. We aim to assess how plausible these scenarios are.
   }
   {We calculate interior and evolution models of Jupiter with such inverted Z-gradient  and use constraints on the stability and the formation of an inverted Z-gradient.}
   {We find that an inverted Z-gradient  at the location of helium rain cannot work as it requires a late accretion and of too much material. We find interior models with an inverted Z-gradient  at upper levels, due to a radiative zone preventing downward mixing, that could satisfy the present gravity field of the planet.
   However, our evolution models suggest that this second scenario might not be in place.}
   {An inverted Z-gradient in Jupiter could be stable. Yet, its presence either at the Mbar level or kbar level is rather unlikely. 
   }


   \maketitle
%
\section{Introduction}

Models of Jupiter's interior, based on Juno gravity data \citep{durante2020}, struggle to agree with measurements of the atmospheric composition \citep{li2020,wong2004,mahaffy2020}. So far, interior models succeeded to bridge the gap, not without difficulty, by relying on different assumptions: by assuming, not comfortably, a higher entropy in the interior or by modifying the equation of state (EOS) \citep{nettelmann2021,miguel2022,howard2023_interior}, by optimising the wind profile \citep{militzer2022} or by including a decrease of the heavy element abundance $Z$ with depth \citep{debras2019}. The last scenario is described by a so-called inverted Z-gradient. Interior models actually favour a low metallicity in the outer envelope (subsolar or solar) while atmospheric measurements from Galileo and Juno suggest a supersolar abundance of heavy elements (around three times the protosolar value, see e.g. \citet{guillot2022,howard2023_interior}). This raises the question if the composition measured in the atmosphere is representative of the entire molecular envelope of Jupiter \citep{2022Icar..37814937H}. 

We discuss in Sect.~\ref{section:1} the concept of an inverted Z-gradient  and the constraints it brings in terms of stability and external accretion. We then present two scenarios.
First, we assess in Sect.~\ref{section:2} the hypothesis of an inverted Z-gradient  located where helium phase separates, as already proposed by \citet{debras2019}. Second, in Sect.~\ref{section:3}, we present a scenario with a similar inverted Z-gradient  but at upper regions, due to a radiative zone.


\section{Inverted Z-gradient : stability, formation}
\label{section:1}

Interior models of Jupiter aim to match the measured gravitational moments, that depend on the density distribution of the planet (see, e.g., \citet{zharkov1978}). However, the difficulty of these models to satisfy the gravitational moments indicates that they seem too dense, especially in outer regions of the envelope (0.1 -- 1\,Mbar) which have a significant contribution to the gravitational moments. Therefore, an inward-decrease of the heavy element content, in agreement with the supersolar atmospheric measurements but then reduced to solar or subsolar at depth, appears as a promising idea. Such inverted Z-gradient  was proposed by \citet{debras2019}, the latter will be discussed in the next section.

\begin{figure}[h]
   \centering
   \includegraphics[width=\hsize]{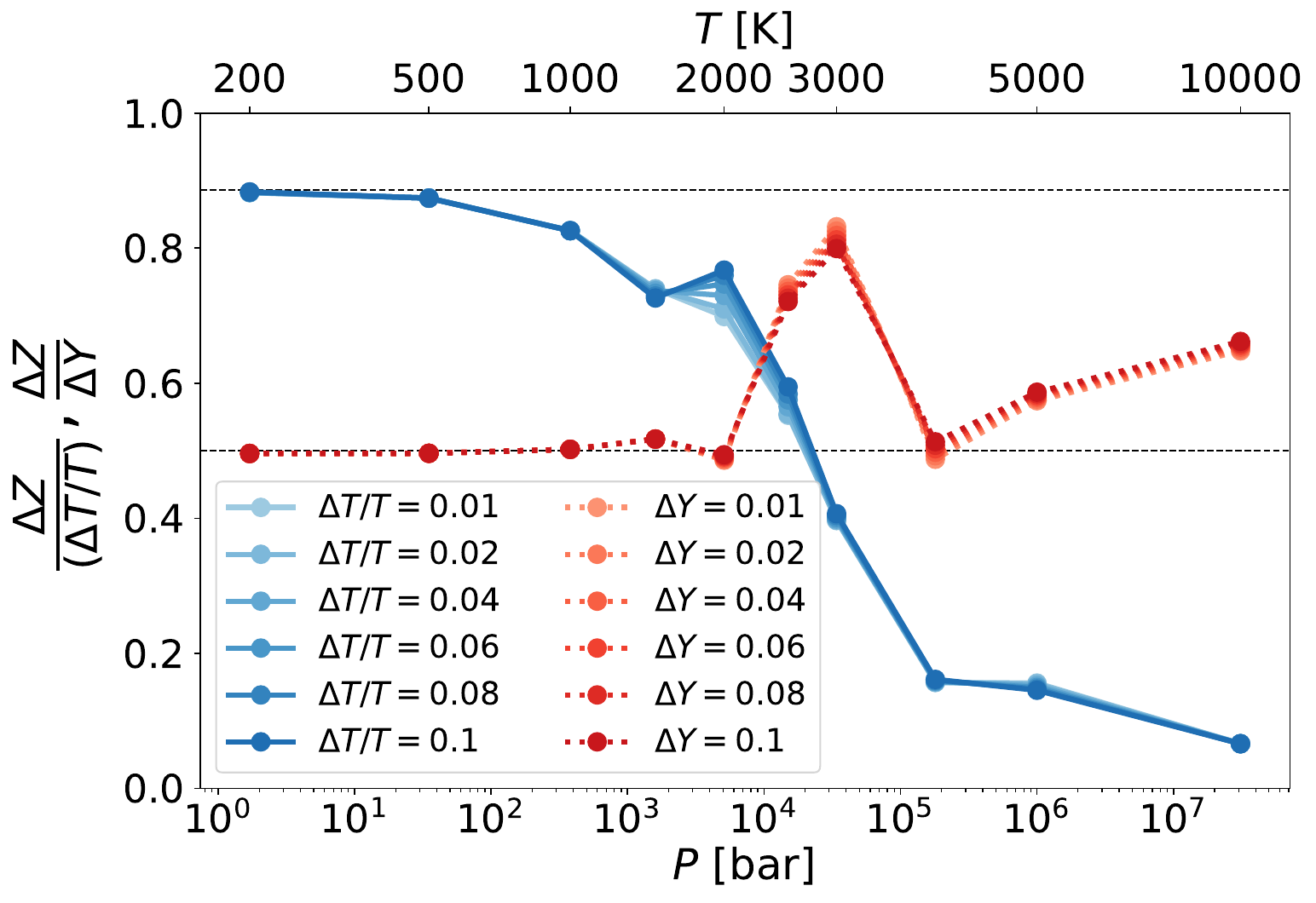}
      \caption{$\Delta Z/ \Delta Y$ and $\Delta Z/(\Delta T/T)$ as a function of pressure and temperature in Jupiter. This shows the maximum increase in heavy elements allowed by increasing $Y$ or by lowering the temperature to ensure stability where the inverted Z-gradient  takes place. The horizontal dashed lines show the values of $\Delta Z/ \Delta Y$ (\textit{bottom}) and $\Delta Z/(\Delta T/T)$ (\textit{top}) for an ideal gas. We stress that the calculation of $\Delta Z$ has here been done using the HG23+CMS19 EOS \citep{chabrier2019,howard2023} for H-He and the SESAME-drysand EOS \citep{sesame1992} for heavy elements. The 1 bar temperature was taken at 170~K.}
         \label{figure:dz_dt}
\end{figure}

Nevertheless, an inverted Z-gradient  requires to be stable against convection to be sustained. It can be balanced either by an increase in the helium mass fraction $Y$ or a decrease in temperature, to make sure that the density $\rho$ still increases with depth. 
We estimate, for Jupiter, the maximum increase in heavy elements that can be afforded by increasing $Y$ or by decreasing temperature. To do so, (i) we calculate $\Delta Z/ \Delta Y$ by equating $\rho(Z_j,Y_j)$ and $\rho(Z_{j+1},Y_{j+1})$ where $\Delta Z=Z_j-Z_{j+1}$ and $\Delta Y=Y_{j+1}-Y_j$ (so that both $\Delta Z$ and $\Delta Y$ have positive values) and (ii) we calculate $\Delta Z/(\Delta T/T)$ by equating $\rho(Z_j,T_j)$ and $\rho(Z_{j+1},T_{j+1})$ where $\Delta T=T_j-T_{j+1}$ is the temperature difference relative to an isentrope. Here, $j$ refers to the layer where the inverted $Z$ gradient takes place. Densities are calculated using the additive volume law and including non-ideal mixing effects \citep{howard2023}:
\begin{equation}
    \frac{1}{\rho(P,T)} = \frac{X}{\rho_{\rm H}(P,T)} + \frac{Y}{\rho_{\rm He}(P,T)} + XYV_{\rm mix} + \frac{Z}{\rho_{\rm Z}(P,T)},
\end{equation}
where $\rho_{\rm H}$, $\rho_{\rm He}$, $\rho_{\rm Z}$ are the densities of hydrogen, helium and heavy elements respectively, $X$, $Y$, $Z$ their respective mass fractions and $V_{\rm mix}$ is the volume of mixing due to hydrogen-helium interactions. Figure~\ref{figure:dz_dt} shows the results. In the ideal gas regime in Jupiter, we expect $\Delta Z/ \Delta Y \sim 0.5$, meaning that the increase in He is required to be at least twice larger than the change in $Z$. We also expect $\Delta Z/(\Delta T/T) \sim 0.9$. It is not exactly 1 because we here assumed a mixture of hydrogen and helium consistent with Galileo's measurement of $Y$ \citep{vonzahn1998}. Using ideal gas relationship $\Delta \mu / \mu = \Delta T / T $ and the definition of the mean molecular weight $\frac{1}{\mu} = \frac{X}{\mu_{\rm H}} + \frac{Y}{\mu_{\rm He}} + \frac{Z}{\mu_{\rm Z}}$, we indeed obtain:
\begin{equation}
    \Delta Z = \frac{\Delta \mu}{\mu} \times \left( \frac{\mu_{\rm Z}\,\mu_{\rm H}}{\mu_{\rm Z}-\mu_{\rm H}}\left( \frac{X}{\mu_{\rm H}}+\frac{Y}{\mu_{\rm He}} \right) \right)
\sim \frac{\Delta T}{T} \times 0.9,
\label{eq:dz_dt_09}
\end{equation}
where $\mu_{\rm H}$, $\mu_{\rm He}$, $\mu_{\rm Z}$ are the molecular weights of hydrogen, helium and heavy elements respectively. The ideal gas regime extends down to the $\sim \,$kbar level. Deeper, non-ideal effects kick in and for instance a bigger decrease in temperature is required to allow an inverted Z-gradient  at deeper regions. One can hence know how much $Z$ can be balanced by an increase in $Y$ or a decrease in temperature, at different levels in Jupiter. 

An inverted Z-gradient  can be stabilised, but vertical transport of heavy material through the stable region may still occur  during the lifetime of Jupiter. We know, for example, that breaking gravity waves \citep{dornbrack1998} and Kelvin-Helmholtz instabilities in the Earth’s stratosphere produce an eddy diffusion coefficient of $10^3~\rm cm^2s^{-1}$ \citep{massie1981}. We assume an eddy diffusion coefficient $K_{\rm zz}$ of $1~\rm cm^2s^{-1}$ which is three order of magnitude smaller, but two orders of magnitude larger than the lower bound, molecular diffusivity. In the case of the presence of a radiative zone (discussed in Sect.~\ref{section:3}), we consider a thickness $L$ of 1000~km for the stable layer. We obtain a diffusion timescale of 
\begin{equation}
    \tau_{\rm mix} \sim 320\, \rm Myr \left( \frac{1\, \rm cm^2s^{-1}}{\textit{K}_{\rm zz}}\right)\left( \frac{\textit{L}}{1000\, \rm km}\right)^2.
\end{equation}
A large uncertainty exists on the eddy diffusion coefficient as well as on the thickness of the stable layer, but maintaining this inverted Z-gradient  on a billion year timescale is rather challenging. In the case of an inverted Z-gradient  located where He phase separates (discussed in Sect.~\ref{section:2}), the thickness of the stable region may be larger, increasing the diffusion timescale to the order of one to ten billion years. 

\begin{figure}[h]
   \centering
   \includegraphics[width=\hsize]{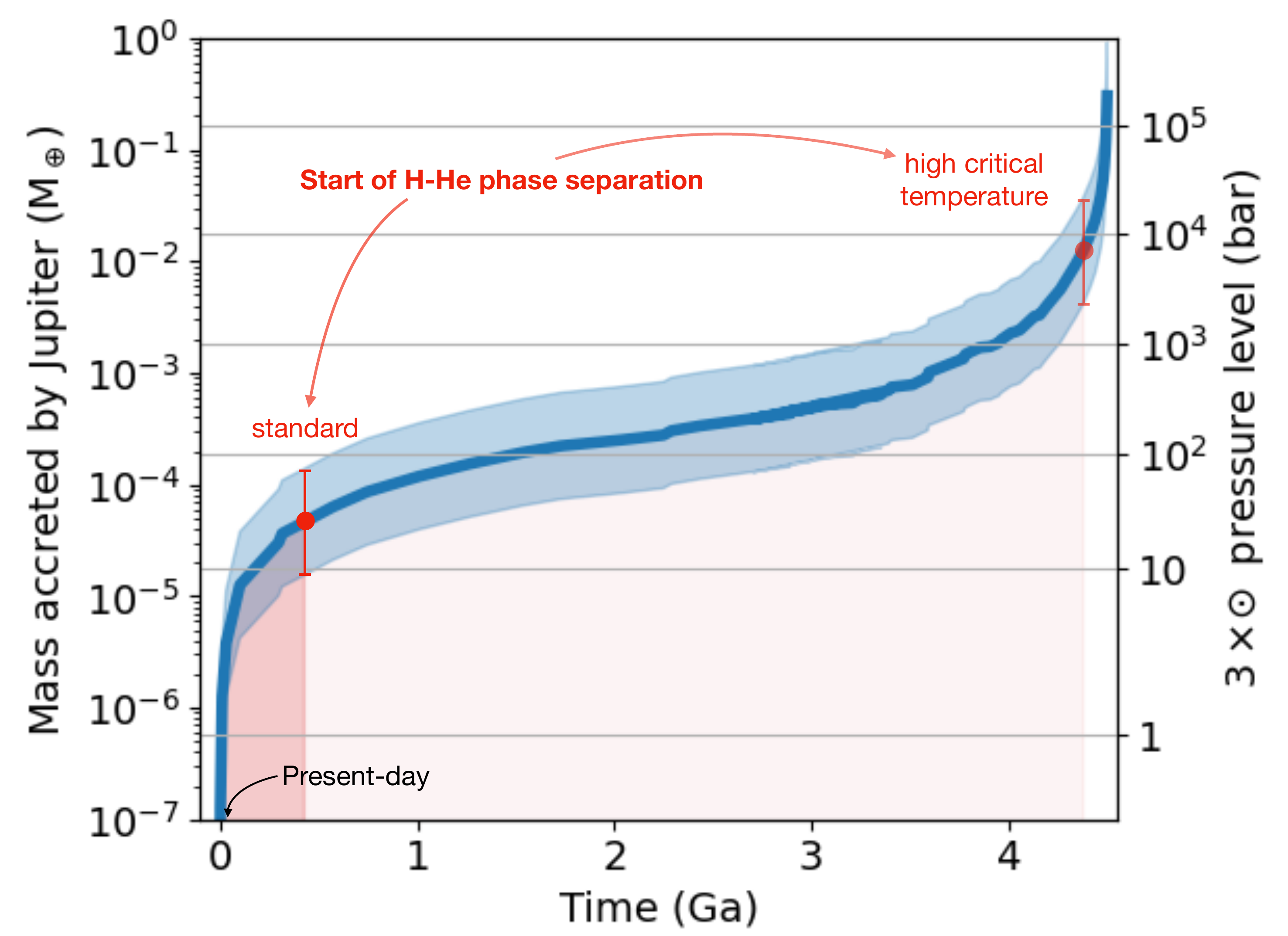}
      \caption{Accretion of material on Jupiter as a function of time from the present (0) to Jupiter's formation 4.5 Ga ago, based on \citet{bottke2023}. In the last one billion years, Jupiter accreted only about $10^{-4}\,M_\oplus$ of material. The right-hand y axis indicates the pressure levels within Jupiter at which the accretion of material would result in a region above this pressure threshold where $Z$ is three times solar. The time at which H-He phase separation started according either to standard models \citep[e.g.,][]{schottler2018_prl} or to experiments indicating a high critical temperature \citep{brygoo2021} are shown in red. The corresponding error bars correspond to the mass accreted since that time.}
         \label{figure:MorbyPlot}
\end{figure}

Furthermore, this inverted Z-gradient  implies some constraints on its origin. First, an enrichment from below is ruled out as internal mixing will tend to homogenise the envelope \citep{vazan2018,muller2020,muller2023}. The enrichment hence needs to be external in order to establish an inverted Z-gradient . We discuss two important aspects of this external enrichment: the amount and the properties of the accreted material. We show on Fig.~\ref{figure:MorbyPlot} how much material can be accreted on Jupiter, through impacts from the destabilised population of the primordial Kuiper belt \citep{bottke2023}. The estimate of the collisional history was based on constraints derived from the craters found on giant planet satellites and the size-frequency distribution of the Jupiter Trojans. The figure also shows the pressure level in Jupiter at which the accretion of such amount of material would lead to a region above this level where $Z$ is three times solar. For instance, in the first 500~Myr after Jupiter's formation, about $2 \times 10^{-3}~M_{\oplus}$ can be accreted, which can lead to a threefold enhancement relative to solar from the top of the atmosphere down to 1~kbar. The two scenarios presented in the following sections will have to satisfy this constraint on the possible amount of accreted material. The occurrence of impacts (large impact or cumulative small impacts) to form an inverted Z-gradient and an investigation of the stability of this region over billions of years is presented in \citet{muller2023}.

\begin{figure}[h]
   \centering
   \includegraphics[width=\hsize]{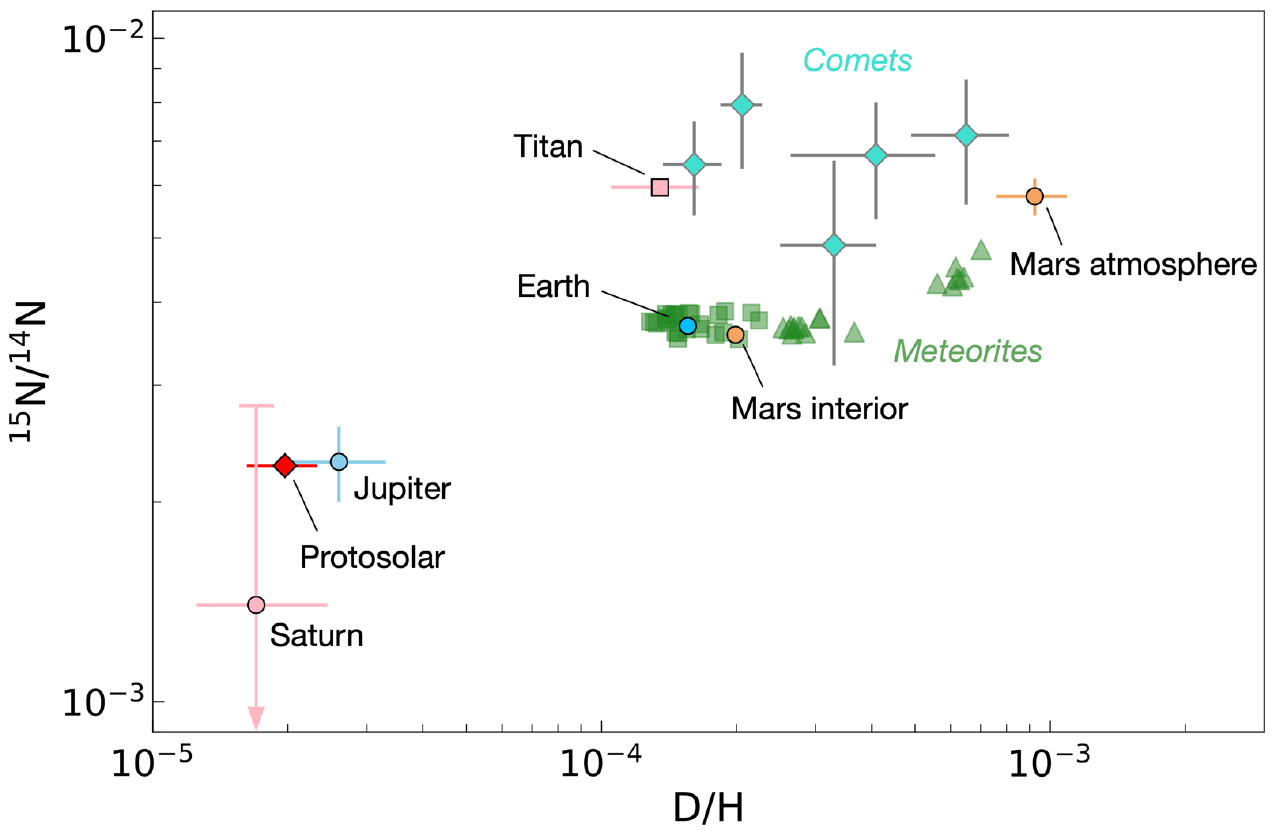}
      \caption{Isotopic ratios of $^{15}\rm N/^{14}\rm N$ and D/H among objects from the solar system. Adapted and updated from \citet{marty2012} and \citet{furi2015}. The D/H of the interior of Mars is a lower limit. The $^{15}\rm N/^{14}\rm N$ value of Saturn is only an upper limit, the central point has been chosen at half that value for illustration purposes. Additional details and used references can be found in Appendix~\ref{appendix:ref_isotopes}.}
         \label{figure:isotopes}
\end{figure}

Finally, we show in Fig.~\ref{figure:isotopes} the isotopic ratios of $^{15}\rm N/^{14}\rm N$ and D/H for objects of the solar system. Only Jupiter (and Saturn) exhibits a protosolar composition of $^{15}\rm N/^{14}\rm N$ \citep{guillot2022} and D/H while all other present objects have supersolar isotopic ratios. Hence, without an early establishment of the inverted Z-gradient, Jupiter's enrichment in heavy elements would have resulted from objects with significantly different isotopic compositions. We would expect the characteristics of a late accretion of heavy elements to match those of objects still present in the solar system today. This constitutes an additional constraint, on the accreted material properties. 
The argument about isotopes is however valid only if we consider that the formed inverted Z-gradient  involves nitrogen. There is a possibility where the accreted material mostly brought carbon and not nitrogen. In fact, the C/N ratio in comet 67P/Churyumov–Gerasimenko is about 29 \citep{fray2017} on average from dust particles, implying a deficit in nitrogen. Combining this value to gas phase measurements, \citet{rubin2019} found a C/N ratio of 22 and 26 in 67P considering a dust-to-ice ratio of 1 or 3. The representativity of this C/N ratio value among other comets is an ongoing research area. The C/N ratio of 67P is in line with 1P/Halley \citep{jessberger1988} and the lower range of 81P/Wild 2 \citep{degregorio2011}. It is also compatible with the chondritic value but large variations are observed in ultracarbonaceous Antarctic micrometeorites (see \citet{engrand2023}). We also stress that the $\rm N_2$ content of comets may have been lost over time. Yet, if the accreted material was depleted in nitrogen, explaining the formation of an inverted Z-gradient  would then require to invoke different processes for different components and lead to a C/N value in Jupiter's atmosphere that is close to protosolar ($\sim \,4.3$, see Table 2 in \citet{guillot2022}). However, we note that uncertainties remain on the composition of the accreted material in both the gaseous and solid phase as well as on the accretion rates.

\section{An inverted Z-gradient  at the helium rain location}
  \label{section:2}

One of the only attempts that succeeded to yield a supersolar abundance of heavy elements in the atmosphere ($Z=0.02$) was from \citet{debras2019}. They included a decrease of heavy elements with depth near He phase separation. It helped reconciling Juno's measurements as it led to lower values of $|J_4|$ and $|J_6|$. To ensure that denser material does not lie on top of lighter material, the decrease in $Z$ was balanced by an increase in $Y$. \citet{debras2019} set the phase separation around 0.1~Mbar in their models, where $\Delta Z \sim 0.015$ and $\Delta Y$ is between $\sim [0.02-0.05]$. Those models hence have $\Delta Z$ between $\sim [0.3-0.75] \times \Delta Y$, ensuring fairly well stability as Fig.~\ref{figure:dz_dt} shows that $\Delta Z$ needs to be smaller than $\sim 0.6 \times \Delta Y$ at this pressure level.

\begin{figure}[h]
   \centering
   \includegraphics[width=\hsize]{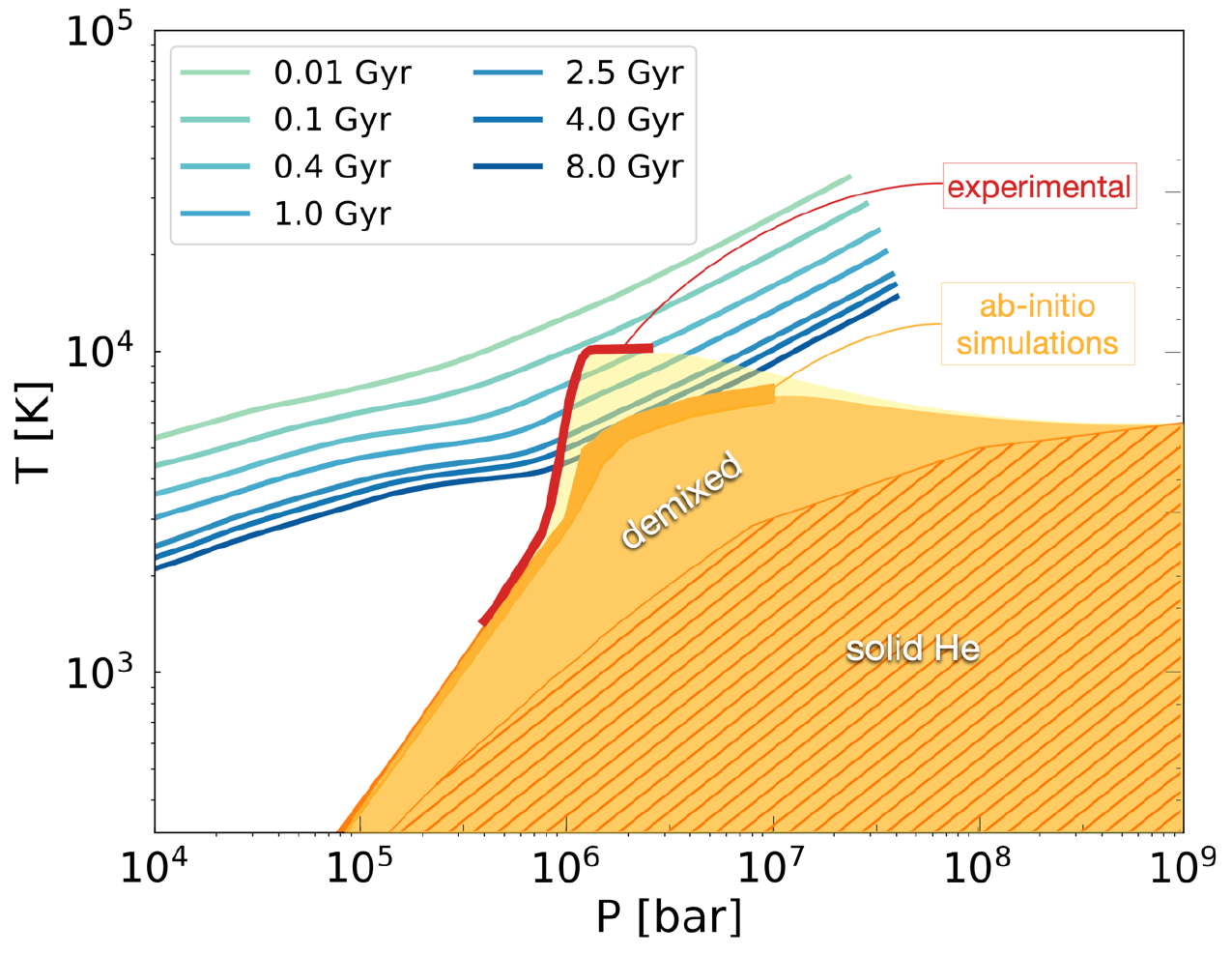}
      \caption{A sequence in evolutionary time of Jupiter interior profiles (from 10~Myr to 8~Gyr) superimposed with a miscibility diagram of H-He. We show the immiscibility curve of experiments from \citet{brygoo2021} (red) and of ab-initio simulations from \citet{schottler2018_prl} (orange). The hashed region is where He becomes solid.}
         \label{figure:phase_diagram}
\end{figure}

However, one of the main explanations of this inverted Z-gradient  is a late accretion of heavy material. This scenario requires such accretion to happen after He demixing occurred in Jupiter, so that the accreted material remains above the location of helium rain. Fig.~\ref{figure:MorbyPlot} shows that to obtain $Z = 3 \times \odot$ above 0.1~Mbar, an accretion of $\sim 0.15\,M_{\oplus}$ of heavy elements is required. Accreting this amount of material is more likely to occur during the early phases of the solar system evolution. More realistic values of the pressure at which He phase separates (a few Mbar \citep{morales2013,schottler2018_prl}) indicate that a few $M_{\oplus}$ of heavy elements are needed to be accreted, which could not be explained. But the timing of the scenario put forward by \citet{debras2019} is challenging. We ran simple evolutionary models of Jupiter (with $M_{\rm core}=10~M_{\oplus}$ and a homogeneous envelope of solar composition). The results are shown in Fig.~\ref{figure:phase_diagram}. We find that He phase separation is expected to occur late in the evolution of Jupiter, i.e. at 4~Gyr (consistent with \citet{mankovich2020}) according to the immiscibility curve of \citet{schottler2018_prl}. But it could occur after 100~Myr at the earliest, if we consider the experimental immiscibility curve from \citet{brygoo2021}. In any case, He demixing is happening relatively late. Such late accretion could not bring more than about $0.01\,M_{\oplus}$ of heavy material (see Fig.~\ref{figure:MorbyPlot}) and would lead to an enrichment with the wrong isotopic composition as discussed in Sect.~\ref{section:1}.

To salvage this scenario, a giant impact between Jupiter and a Mars-mass object could be envisioned. Such objects are not found in the current Kuiper belt population. Evaluating the likelihood of such an impact at different ages is crucial. Based on conventional H-He phase diagrams, this impact should have occurred very late, in the last 500~Myr, making it an extremely low-probability event. It might also have occurred earlier, as suggested by the high critical demixing temperature of \citet{brygoo2021}. In both cases, however, the impactor should have brought little nitrogen or had a different composition from the observed small objects in the solar system.

\section{An inverted Z-gradient  at uppermost regions, due to a radiative zone}
  \label{section:3}

We now envision an inverted Z-gradient located at upper regions ($\sim \,$kbar) and established early (less than 10~Myr). Our hypothesis is that the presence of a radiative zone prevents downward mixing. A radiative region could exist in the upper envelope, as suggested by \citet{guillot1994} (between 1200 and 2900~K) and by \citet{cavalie2023} (between 1400 and 2200~K). A depletion of alkali metals would bring support to the existence of a radiative layer \citep{bhattacharya2023}. Accreted heavy material on top of this radiative zone may thus be prevented from mixing with the rest of the envelope below this radiative zone. We note that the presence of a radiative zone is a separate question of getting a higher $Z$ value above this radiative layer, which is what we focus on here.

\begin{figure}[h]
   \centering
   \includegraphics[width=\hsize]{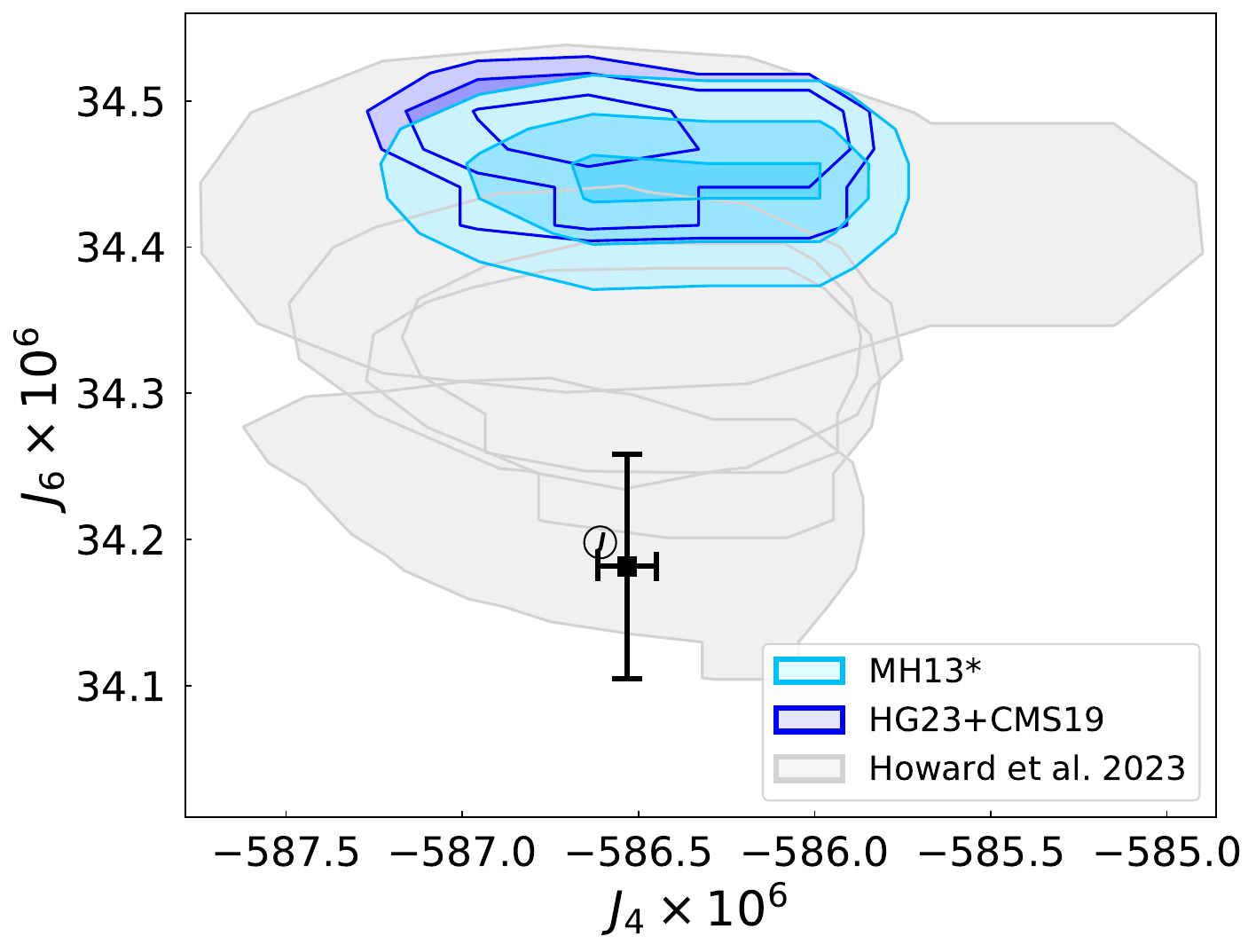}
      \caption{$J_6$ vs. $J_4$ of models including a radiative zone ($Z$ being $3\, \times$ protosolar above), using the MH13* \citep{militzer2013} or HG23+CMS19 \citep{chabrier2019,howard2023} EOS. These models are compared to results by \citet{howard2023_interior} with no radiative zone nor inverted Z-gradient ($Z$ being only $1.3\, \times$ protosolar in the atmosphere). The full co-variations of the MCMC calculation using MH13* are given in Appendix~\ref{appendix:cornerplot}. The circled J shows the Juno measurements \citep{durante2020} and the error bar accounts for differential rotation, as in \citet{miguel2022}.
      }
         \label{figure:j6j4}
\end{figure}

First we ask: Can we find interior models with such radiative zone that satisfy the present gravity field measured by Juno? To answer this question, we use the opacities from \citet{guillot1994} (including absorption by $\rm H_2$, He, $\rm H_2 O$, $\rm CH_4$, $\rm NH_3$) to set a radiative region and implement an inverted Z-gradient.
Fig.~\ref{figure:dz_dt} shows that an inverted Z-gradient  can be stabilised by a sub-adiabatic temperature gradient. Around the $\sim \,$kbar level, $\Delta Z$ needs to be smaller than about $0.7 \times \Delta T/T$. Here, our models have $\Delta T/T$ of about 10\%, allowing an increase of $Z$ of three times the protosolar value ($Z \sim 5\%$, \citep{asplund2021}) and hence ensuring stability. We ran Markov chain Monte Carlo (MCMC) calculations (as in \citet{miguel2022,howard2023_interior}).  At first, we could not find models fitting the equatorial radius and the gravitational moments of Jupiter. In this case, the radiative zone was extending from 1200 to 2100~K. We then parameterised (arbitrarily multiplying by 5) the opacities and could obtain solutions, with a radiative region extending from about 1600 to 2100~K. Thus, the possible location and extent of the radiative zone may be constrained by the gravity data. Fig.~\ref{figure:j6j4} shows the gravitational moments $J_4$ and $J_6$ of these models, for two EOSs (the full posterior distributions are given in Appendix~\ref{appendix:cornerplot} for one EOS).  
We find that interior models including such radiative zone can satisfy the observed gravitational moments from Juno (at $3~\sigma$ in $J_6$) as well as the compositional constraints on the atmosphere.

\begin{figure}[h]
   \centering
   \includegraphics[width=\hsize]{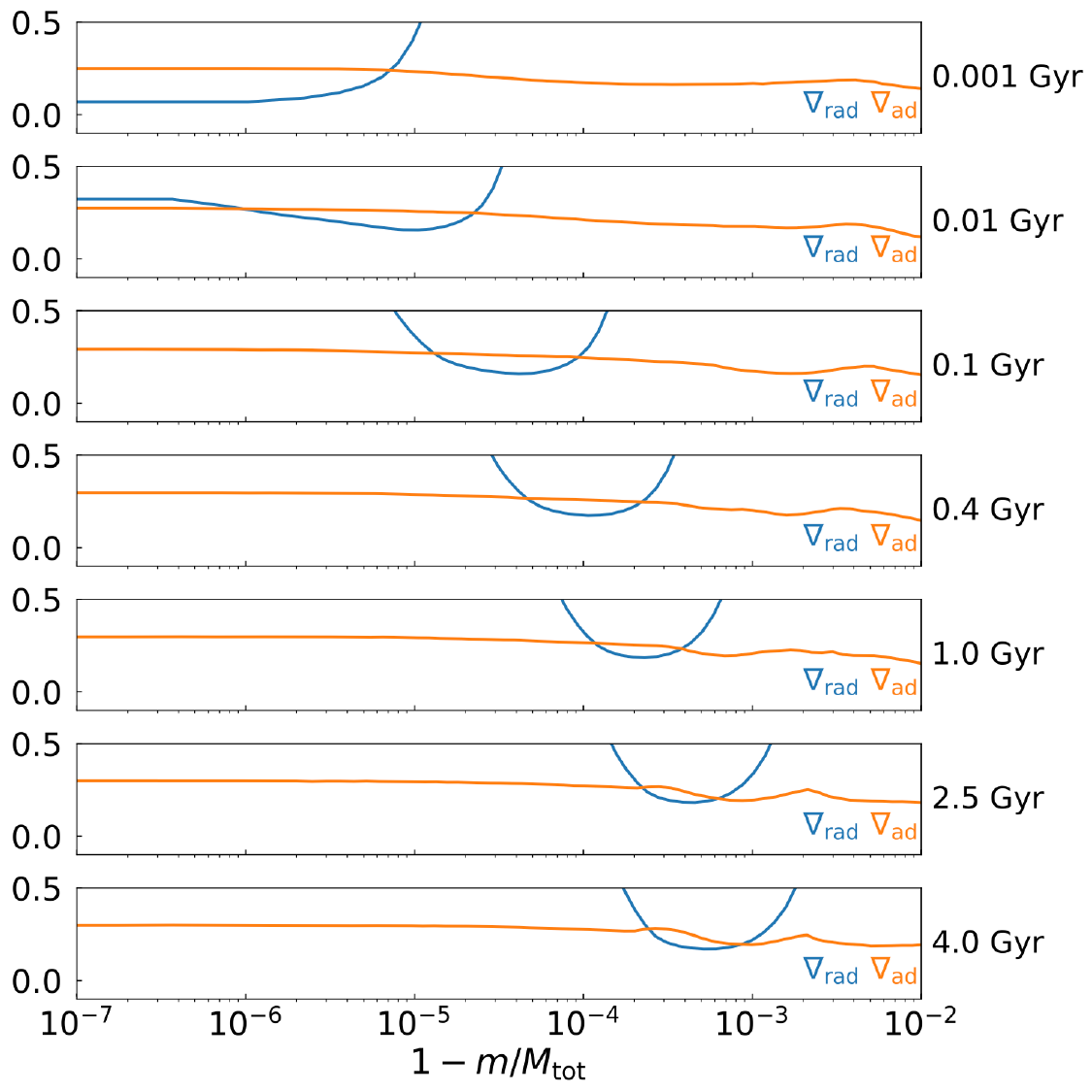}
      \caption{Comparison of the radiative and adiabatic temperature gradients in Jupiter, at ages ranging from 1~Myr to 4~Gyr. We estimate the minimum value of $\nabla_{\rm rad}$ as the approximate upper limit of the radiative region. The evolution models consist of a central core of $10~M_{\oplus}$ and a homogeneous envelope of solar composition.}
         \label{figure:evol}
\end{figure}

To get $Z = 3 \times \odot$ above 1~kbar, an accretion of $\sim 2 \times 10^{-3}\,M_{\oplus}$ of heavy elements is required, which can be done in the first few hundred million years (see Fig.~\ref{figure:MorbyPlot}). The isotopic constraints mentioned in Sect.~\ref{section:1} imply that a late delivery of heavy material can hardly be possible. Hence, we examine whether the inverted Z-gradient  could have been formed early and maintained in Jupiter. To this end, we use again our evolutionary models presented in Sect.~\ref{section:2} and include now a radiative zone using the parameterised opacities. Figure~\ref{figure:evol} compares the radiative and adiabatic temperature gradients, from 1~Myr to 4~Gyr. The radiative zone is located roughly where the radiative gradient is lower than the adiabatic gradient. The radiative region appears around 10~Myr and is progressively shifted to deeper regions. Thus, the initially enriched material above the radiative zone will progressively mix with material of protosolar composition as the radiative zone is shifted to deeper levels. Such behaviour of the radiative zone was already predicted by \citet{guillot1999}. Considering that the mass above the radiative region at 10~Myr is $\sim 10^{-5}~M_{\rm J}$ and increases to $\sim 3.10^{-4}~M_{\rm J}$ at 4~Gyr, the Z-gradient  at 10~Myr must be high enough so that the abundance of heavy elements becomes approximately three times the protosolar value nowadays. If the disk phase does not exceed 10~Myr, a $Z$ value of 60 times the protosolar value ($\Delta Z \sim 0.9$) is required above the radiative zone, at 10~Myr. Keeping such a $\Delta Z$ and ensuring stability may be hard since a significant $\Delta T/T$ would be required to prevent mixing (the factor of 0.9 in Eq.~\ref{eq:dz_dt_09} would even be lower given the increased molecular weight due to a much higher $Z$ value), making the scenario rather unlikely. Furthermore, diffusion through the radiative zone is expected after a few hundred million years (see Sect.~\ref{section:1}), making the scenario even more challenging. However, this behaviour of the radiative zone as the planet evolves is one case corresponding to the use of a specific opacity table and the details of the evolution code. Further investigation of this scenario is therefore required.

\section{Conclusion}
The inverted Z-gradient  is an appealing idea for interior models to explain both the gravity field and the atmospheric composition of Jupiter. It may also be of interest to Saturn as reconciling interiors models with the measured metallicity is challenging too \citep{mankovich2021}, noting that helium rain is expected to start earlier. An inverted Z-gradient can be stabilised by either an increase in the helium mass fraction or a decrease in temperature. However, as we show here, such Z-gradient in Jupiter at the location of helium rain, as proposed by \citet{debras2019}, is rather unlikely as it requires to accrete an excessive amount of material, that cannot be justified from collisional evolution models of the solar system. It also requires a late accretion, that isotopic constraints do not allow. 
An inverted Z-gradient, established early and at upper regions ($\sim \,$kbar), due to a radiative zone, might be a solution. We show that such a scenario works from the point of view of the present gravity data and enough material may be accreted. Nevertheless, this radiative zone appears around 10~Myr and is shifted to deeper regions with time. 
Such inward-shift of the radiative zone requires at $\sim 10~$Myr a significant Z-gradient  ($\Delta Z \sim 0.9$ in our case), that is hard to be stabilised.
However, our calculations rely on a specific (and parameterised) opacity table used here. Updated opacity data (see, e.g. \citet{tennyson2012,freedman2014}) could produce a radiative region at a different location and with a different evolution, changing the required mass that needs to be accreted to enrich the outer envelope. Furthermore, despite this work being based on the latest considerations regarding the quantity and properties of the materials enriching the atmosphere, our knowledge of Jupiter's potential accreted material is still incomplete. Yet, in our setup, the hypothesis of a radiative zone that prevents downward mixing is rather unlikely. Alternative scenarios such as an inverted gradient of helium instead of heavy elements as well as further investigation of the topic are required to resolve Jupiter's metallicity puzzle.


\begin{acknowledgements}
We thank A. Morbidelli for his precious input on constraints on Jupiter's accreted mass and D. Bockelée-Morvan for sharing valuable references about comet composition. We thank the Juno Interior Working Group for useful discussions. This research was carried out at the Observatoire de la Côte d'Azur under the sponsorship of the Centre National d'Etudes Spatiales.
\end{acknowledgements}

%
   \bibliographystyle{aa} 
   \bibliography{aanda} 
%

\onecolumn
\begin{appendix} 
\section{Corner plot of models using MH13* and including a radiative zone}
\label{appendix:cornerplot}
\begin{figure*}[h]
   \resizebox{\hsize}{!}
            {\includegraphics{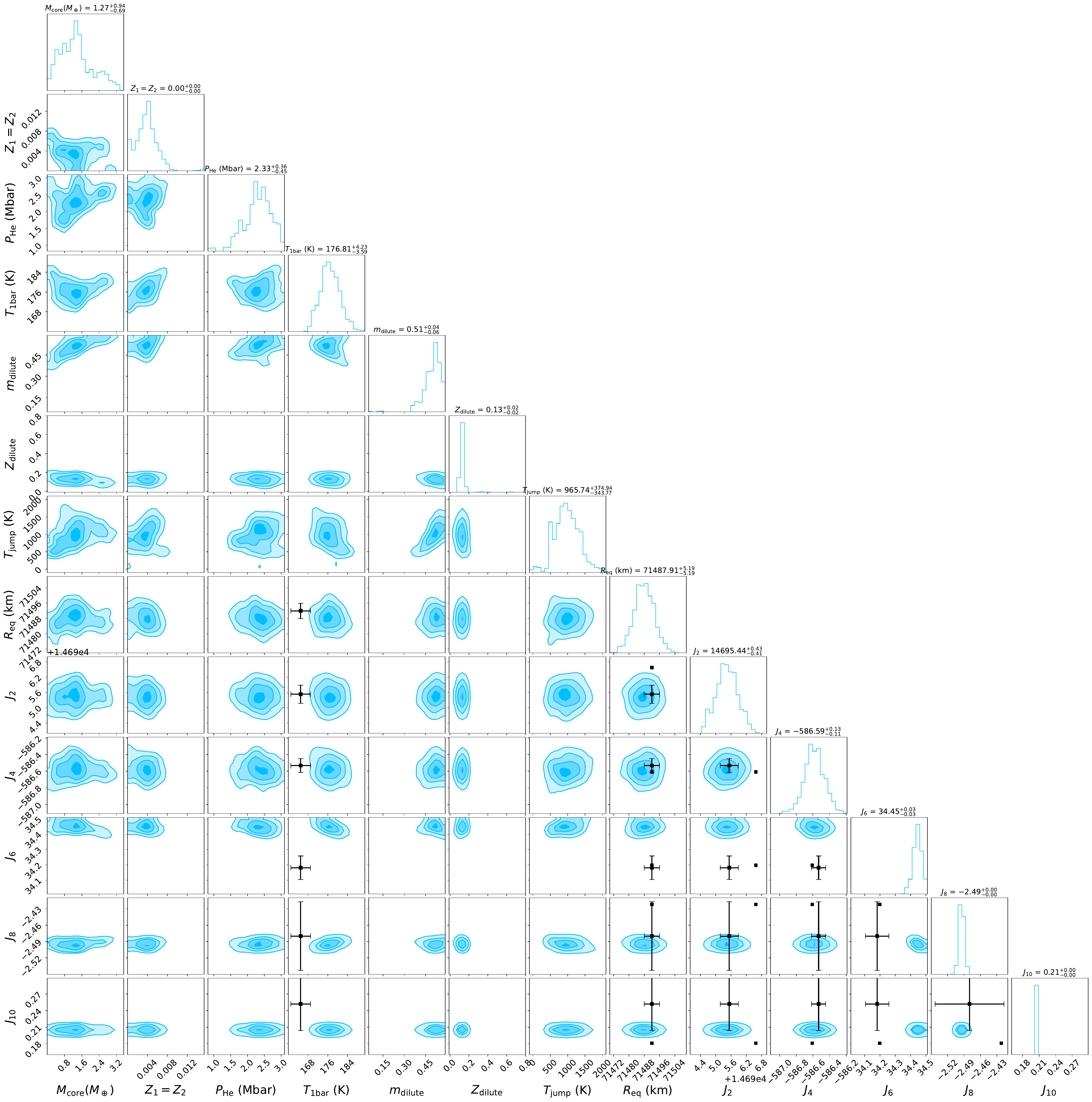}}
      \caption{Posterior distributions obtained with the MH13* EOS \citep{militzer2013}, including a radiative zone (opacities multiplied by 5). The black points correspond to the measured $J_{2n}$ by Juno. The black error bars correspond to Juno's measurements accounting for differential rotation for the $J_{2n}$ and Galileo's measurement for $T_{\rm 1bar}$.
              }
         \label{appendix:figure_cornerplot}
\end{figure*}

Figure~\ref{appendix:figure_cornerplot} shows the posterior distributions of the MCMC simulations using the MH13* EOS \citep{militzer2013}, including a radiative zone and an inverted Z-gradient .

\clearpage
\newpage

\section{Additional details and references for isotopic ratios of $^{15}\rm N/^{14}\rm N$ and D/H, used in Fig. 3}
\label{appendix:ref_isotopes}
This appendix lists the references used to plot the isotopic ratios shown on Fig.~\ref{figure:isotopes}. Jupiter's data are coming from Galileo \citep{mahaffy1998,owen2001}: $\rm D/H=(2.6 \pm 0.7)\times 10^{-5}$ and $^{15}\rm N/^{14}\rm N=(2.3 \pm 0.3)\times 10^{-3}$. Protosolar values are from \citet{marty2011,geiss2003}. Earth's data are from \citet{anders1989,michael1988}. Mars' data are from \citet{mathew2001} and \citet{wong2013,webster2013} respectively for its interior and its atmosphere. The D/H of the interior is a lower limit as large variations are measured in martian meteorites \citep{saito2020}. Saturn's data are from \citet{lellouch2001,fletcher2014}: $\rm D/H=1.7^{+0.75}_{-0.45}\times 10^{-5}$ and $^{15}\rm N/^{14}\rm N=2.8\times 10^{-3}$. The $^{15}\rm N/^{14}\rm N$ value is an upper limit. Titan's data are from \citet{niemann2010,abbas2010}. For meteorites, bulk isotopic ratios (\textit{squares}) and values in insoluble organic matter (IOM) (\textit{triangles}) are displayed. Here are shown data for various types of chondrites (CI, CM, CO, CR, CV) from \citet{kerridge1985,aleon2010}. Data for 5 comets (103P/Hartly, C/2009 P1 Garradd, C/1995 O1 Hale-Bopp, 8P/Tuttle and C/2012 F6 Lemmon, from left ro right) are displayed, taken from \citet{manfroid2009,biver2016,shinnaka2016,lis2019} and \citet{bockelee2015} and references therein. We mention that the average $^{15}\rm N/^{14}\rm N$ value for 21 comets has been found to be $0.007 \pm 0.001$ \citep{manfroid2009}.

\end{appendix}

\end{document}